\def \yskip{\penalty-50\vskip3pt plus 3pt minus 2pt}
\def \reference{\par \yskip \noindent \hangindent .4in \hangafter 1}
\def \abc#1#2#3#4 {\reference#1, {\sl#2}, {\bf#3}, #4}
\def \blank {\lower 5pt\hbox to 0.75in{\hrulefill}}

\def \cm{~\rm{cm}}
\def \s{~\rm{s}}
\def \km{~\rm{km}}

\def \K{~\rm{K}}

\def \AU{~\rm{AU}}

\def \yrs{~\rm{yrs}}
\def \yr{~\rm{yr}}
\def \pc{~\rm{pc}}

\def \lesssim{\mathrel{<\kern-1.0em\lower0.9ex\hbox{$\sim$}}}
\def \gtrsim{\mathrel{>\kern-1.0em\lower0.9ex\hbox{$\sim$}}}

\documentstyle[11pt,aasms,tighten,flushrt]{article}

\begin{document}
\small

\setcounter{page}{1}

\begin{center}
\bf
A MODEL FOR THE STRINGS OF $\eta$ CARINAE
\end{center}

\begin{center}
Noam Soker\\
Department of Physics, University of Haifa at Oranim\\
Oranim, Tivon 36006, ISRAEL \\
soker@physics.technion.ac.il 
\end{center}


\begin{center}
\bf ABSTRACT
\end{center}

We propose a model based on ionization shadows to explain the
formation of the long and narrow strings of $\eta$ Carinae.
 Five strings are known, all located along the symmetry
axis outside the Homunculus.
  The model assumes that each string is formed in a shadow behind
a dense clump near the symmetry axis.
 The surrounding gas is ionized first, becomes much hotter, and
compresses the gas in the shadow. 
 This leads to the formation of a radial, dense, long, and narrow
region, i.e., a string. 
 Later the neutral material in the strings is ionized, and becomes
brighter. Still later it re-expands, and we predict that in
$\sim 100-200 \yrs$ the strings will fade.
 The condition for the model to work is that the ionization front,
due to the diffuse ionizing recombination radiation of the
surrounding gas, proceeds into the shadow at a velocity slower
than the compression speed, which is about the sound speed.
 From that we get a condition on the mass loss rate of
the mass loss episode that formed the strings, which
reads $\dot M_s \lesssim 10^{-4} M_\odot \yr^{-1}$.
 The model can also explain the strings in the planetary nebula
NGC 6543.


\clearpage

\section{INTRODUCTION}

 There are several open questions regarding the formation of the
nebulosity around the massive star $\eta$ Carinae
(Davidson \& Humphreys 1997; hereafter DH97).
 Some of the questions concern the nature and formation process of
the strings (Weis, Duschl \& Chu 1999, hereafter WDC;  Weis 2001).
 WDC identified 5 strings, which are long, $\sim 0.1 \pc$, and narrow,
width of $\sim 0.002 \pc$, almost straight filaments, denser than
their environment, and located close to the symmetry (major) axis of
the Homunculus, but outside the Homunculus.
 They expand radially, following a Hubble-type law, with velocity
increasing from $\sim 450 \km \s^{-1}$ in the parts closest to the
central star, to $\sim 900 \km \s^{-1}$ at their ends away from
the central star (WDC; Weis 2001).
 From their kinematics it seems that the material in the strings was
expelled during the Great Eruption of 1850, or just prior or after the
eruption (ADS).
 Interestingly, the planetary nebula (PN) NGC 6543 has filaments
similar in many properties to those of $\eta$ Car (WDC);
they have the same general shape, same relative location, i.e.,
outside the main nebula near the symmetry (major) axis, and
they are also much fainter than the main shell.
 Kinematically, they expand at much lower velocities
$\sim 50 \km \s^{-1}$ (Balick \& Preston 1987), as expected
in PNe, and their distance to the central star of NGC 6543 is about
half the distance of the filaments of $\eta$ Car to the center.
 As discussed later, the proposed model can also explain the
formation of the filaments in the PN NGC 6543.

 In the present paper we propose a model for the formation of
the strings.
 The explanation for the Hubble-type expansion law of the strings
is beyond its scope. 
 We only propose a model for the formation of the long and
narrow radial strings, assuming they are formed from material that
is already expanding by a Hubble-type law.
 We propose that each string was formed by the compression of
neutral cool gas in an ionization shadow.
 The surrounding gas was ionized first, and its temperature and pressure
became larger by a factor of $\sim 10$ than the still cool gas in the
shadow.
 Hence the surrounding gas compresses the gas in the shadow, forming a
radially long and narrow tail in the shadow, a string. 
 Nor do we deal with the origin of the material, which could
have been expelled by $\eta$ Car itself, or was first accreted
by a companion and then expelled by the companion (Soker 2001).
 The proposed model is outlined in section 2.
 In section 3 we show that it is quite possible that the companion
(if it exists) also played a role in the ionization process.
 The conditions for the formation of the strings from ionization shadows
are derived in section 4.
 A summary of main results is in section 5.  

\section{THE PROPOSED MODEL}

 The proposed model for the formation of the strings is that
of an ionization shadow behind a dense clump.
 The ionization front proceeds much slower inside the dense clump,
hence the clump shades the region behind it from the ionizing radiation,
at least during part of the time until it is completely evaporated
(e.g., Lopez-Martin {\it et al.} 2001).
 The shadow stays cool, at $T < 10^3 \K$, while its surroundings are
ionized and heated, hence it is compressed to a higher density
(Canto {\it et al.} 1998).
 An ionization shadow model for the formation of radially
aligned narrow structures was discussed in the context of
PNe (Soker 1998, 2000).
 The strings of $\eta$ Car are different in several respects
 from the situations studied by Soker (1998, 2000).
 First, the strings in $\eta$ Car are very narrow, hence
a steady-state neutral tail behind a dense clump cannot be
reached (Canto {\it et al.} fig. 4.).
 Second, the structures studied by Soker were located in a dense region
in the PN shell, unlike the strings of $\eta$ Car, which are located
in relatively low density regions outside the main shell of the nebula.
 The very narrow strings of the PN NGC 6543 (WDC) are also located
in a (relatively) low density region outside the dense shell of NGC 6543.
 Third, the material in $\eta$ Car expands with velocities of
$\gtrsim 400 \km \s^{-1}$, much faster than the sound speed,
unlike in PNe, where the two speeds are comparable.

 Because of these differences, the proposed model does not deal with
a steady-state model, where a neutral shadow exists until the
shadowing clump is ionized; instead it is a dynamic model, where
the strings exist for a relatively short time.
 Before describing the model, we emphasize that the relevant mass
loss rates are those along the polar directions
(mass loss rate per unit solid angle) were the strings are found,
and not the average or equatorial mass  loss rates.
 The scenario has four main phases.
\newline
{\bf Phase 1 (20 years):}
During the 20 years of the Great Eruption of 1850 the star lost
$\sim 2.5 M_\odot$ (DH97).
 Because of the large optical depth during the eruption, the gas
in the wind recombined and cools, down to $T < 10^3 \K$,
after leaving the star and expanding.
 From the shape of the nebula, it seems
that the mass loss rate along the polar directions was lower than
average, allowing earlier ionization along these directions at 
later times.
\newline
{\bf Phase 2: ($\sim 50$ years):}
After the eruption, the average mass loss rate probably droped to
something like its present value of
$\dot M \sim 10^{-4} M_\odot \yr^{-1}$ (DH97; Corcoran {\it et al.} 2001b).
 The mass loss rate increased for a short period of time during
the lesser eruption of 1890, but we ignore this mass loss episode
(we have no data on the polar mass loss rate, which is 
relevant for us).
 As the dense material from the great eruption expanded, the ionization
front moved outward.
 With ionization of fresh neutral material being disregarded, the
ionization front $r_i$ is given by the equality of the photon luminosity
$\dot S_1$ to the total recombination rate
\begin{equation}
\dot N = \int_{r_m}^{r_i} 4 \pi \alpha n_e n_i r^2 dr,
\end{equation}
where $r_m$ is the minimum radius outer to which recombination
occurs, $\alpha$ is the recombination coefficient, and $n_e$ and
$n_i$ are the electron and ion density, respectively 
(for the purpose of the present paper we can take $n_i=n_p$,
where $n_p$ is the proton number density).
 The wind's density is $\rho = \dot M_{GE} / (4 \pi r^2 v)$,
where $\dot M_{\rm GE}$ is the mass loss rate along
the polar directions during the great eruption and $v$ the expansion
velocity.
 We first neglect recombination by the lower density wind blown during
the second phase (see next section), and consider the dense material from
the Great Eruption, with its inner boundary expanding at
$v \sim 650 \km \s^{-1}$, and its outer boundary ejected 20 years
earlier (the duration of the Great Eruption).
 Substituting typical values, taking the mass loss rate along
the polar direction to be half the average value,
and integrating equation (1), we obtain the time when the the entire
nebula along the polar direction was ionized.
\begin{equation}
t \sim 30 
\left( {{\dot S_1}\over{10^{50} \s^{-1}}} \right)^{-1}
\left( {{\dot M_{\rm GE}}\over{0.05 M_\odot \yr^{-1}}} \right)^2
\left( {{t + 20 \yrs}\over{50 \yrs}} \right)^{-1}
\yr
\end{equation}
 We note that a small drop in the mass loss rate along the
polar directions can shorten this phase.
 On the other hand, including ionization of the wind blown by
the central star during this phase will give a longer time.
 Hence we estimate this phase to last $\sim 50 \yrs$.
 The possibility that the companion plays a role in the ionization process
is examined in the next section.
\newline
{\bf Phase 3 ($\sim 100-200$ years):}
 The models assumes that very dense clumps exist along and near
the polar directions.
 These can be formed from in situ instabilities, or from
inhomogeneous mass loss process.
Their formation process is beyond the scope of the present paper.
 The dense clumps form a shadow behind them, so that while the
surroundings of the shadow are ionized by stellar radiation, the
shadow is ionized by the diffuse recombination radiation.
 If the ionization, and heating, of the shadow proceeds at a time scale
longer than the sound crossing time of the shadow, the hot surrounding
gas compresses the cooler gas in the shadow to a higher density,
forming a long straight string.
 The conditions for this to occur are derived in section 4.
 A typical string width is $d_s=0.002 \pc$ (Weis 2001).
  The compression proceeds at a velocity of the order of the
sound speed of the surroundings, $\sim 15 \km \s^{-1}$, so
that the sound crossing time of a half width (since the compression
proceeds from all directions) is
\begin{equation}
t_s = 70 \left( {{d_s}\over{0.002 \pc}} \right) \yrs .
\end{equation}
 The initial width of the shadowing clump is larger, but  after
compression (see section 4) the density increases by a factor of
$\sim 10$ (assuming a pressure equilibrium between the two regions) 
and the width of the compressed region decreases by a
factor of up to $\sim 3$.
 This means that the sound crossing time can be longer, and so
can the ionization time by the diffuse radiation.
 Hence this phase may last $\sim 150-200 \yr$.
 If the shadowing clump is ionized in the mean time, then this phase
will be shorter.
Over all, we take this phase to last $100-200 \yr$.
\newline
{\bf Phase 4 ($ \gtrsim 200-300$ years):}
Eventually the shadow is ionized by
the diffuse radiation, grows hotter, and reexpands
at a time scale of $\sim 100$ years.
 If the shadowing clump still exists, it stays somewhat cooler and
denser that its surroundings, by a factor of $\sim 2$
(Canto {\it et al.} 1998), until the entire clump is evaporized and
the stellar radiation reaches the shadow. 
 It is very likely that the reexpansion phase has started already.
We therefore predict that in $\sim 100$ years the strings will fade
substantially, and may even disappear.

\section{PRIMARY VERSUS SECONDARY IONIZATION }

 The proposed model involves ionization by the central source along and
near the polar directions, as discussed in the previous section.
 This section examines the possibility that the companion (if it exists)
plays a role, even a dominant one, in the ionization
process.
 Let the mass loss rate from the primary star ($\eta$ Car itself)
and the expansion velocity during the relevant ionization epoch
be $\dot M_i$, and $v_i$, respectively, so that the wind's density
in a spherical geometry is $\rho = \dot M_i / (4 \pi r^2 v_i)$.
 Neglecting ionization of new material, assuming spherical symmetry,
and neglecting material from the Great Eruption, we find that
the ionization front $r_i$ is given by the equality of the photon
luminosity $\dot S_1$ with the total recombination rate.
 The recombination rate is given by equation (1), with the mass loss rate
and velocity being $\dot M_i$ and $v_i$, respectively,
and $r_m$ is the stellar radius, $r_m=R_1=0.4AU$ (DH97).
 Substituting other typical values we obtain the condition for ionization
of the entire nebula
\begin{equation}
{{\dot S_1}\over{10^{50} \s^{-1}}} \gtrsim 0.2
\left( {{\dot M_i}\over{10^{-4} M_\odot \yr^{-1}}} \right)^2
\left( {{v_i}\over {500 \km \s^{-1}}} \right)^{-2}
\left( {{R_1}\over {0.4 \AU}} \right)^{-1}.
\end{equation}

 We now consider the ionizing radiation from the secondary, assuming
that during the relevant ionization phase (see next section) the
secondary does not blow a dense neutral wind along the polar directions,
i.e., $\dot M_2 \lesssim 10^{-5} M_\odot \yr^{-1}$. 
We also assume that there is no dense accretion flow along the polar
directions near the companion.
 For the companion and orbital parameters we take values which were
used in recent years (e.g., Ishibashi {\it et al.} 1999;
Damineli {\it et al.} 2000; Corcoran {\it et al.} 2001a, 2001b): 
mass $M_2=30 M_\odot$, semimajor axis $a=15 \AU$ and eccentricity
$e=0.8$. 
  Let the momentary distance between the two stars be $D$, and let
$h$ be the length along a line perpendicular to the
orbital plane and crossing the momentary position of the
secondary.
 The density of the wind blown by the primary as a function of $h$ is
$\rho(h)= \dot M_i / [4 \pi (D^2 + h^2) v_i]$.
 The total recombination rate per unit solid angle along
 the direction $h$ is given by
\begin{equation}
\dot N = \int_{r_m}^{r_i} 4 \pi \alpha n_e(h) n_i(h) h^2 dh.
\end{equation}
 The integral can be solved analytically.
 Taking $r_i \gg r_m=0$, i.e., the entire nebula is ionized,
we find
\begin{equation}
\dot N = 4 \pi \alpha
{{n_e}\over{\rho}} {{n_i}\over{\rho}}
\left( {{\dot M_i}\over{4 \pi v_i}} \right)^2
{{\pi}\over{4 D}}.  
\end{equation}
 Note that along a line from the primary the last term
reads $1/R_1$ instead of $\pi/4 D$.  
 Substituting typical values for the companion:
$\dot S_2$, which is the ionizing photon luminosity for a main
sequence star of $M_2 \simeq 30 M_\odot$, for a semimajor axis of
$15 \AU$ the orbital separation will be $> 15 \AU$ during most
of the orbital period, so that we scale $D$ with $20 \AU$.
 The condition for ionizing the entire material along the polar
directions by the companion reads 
\begin{equation}
{{\dot S_2}\over{2 \times 10^{48} \s^{-1}}} \gtrsim 0.2
\left( {{\dot M_i}\over{10^{-4} M_\odot \yr^{-1}}} \right)^2
\left( {{v_i}\over {500 \km \s^{-1}}} \right)^{-2}
\left( {{D}\over {20 \AU}} \right)^{-1}.
\end{equation}

 By comparing the conditions for full ionization along the polar
directions,  equation (4) for the primary and equation (7) for
the secondary, we see that in principle both can be the ionization
source, during the phases when the mass loss rate is
$\dot M_i \lesssim 10^{-4} M_\odot \yr^{-1}$, the third
and fourth phases in the scenario plotted in the previous section.
 Which of the two ionizing source dominates depends on details such
as the departure of mass loss from sphericity and the properties
of the wind blown by the companion.
 In any case, the proposed model for the strings can hold for the
cases where no binary companion exists.
 Two phenomena suggest that the companion indeed plays a role.
 (1) All the presently known strings are located to the same side of
the long axis of the $\eta$ Car nebula (WDC).
This departure from axisymmetry hints at a role played by a
companion in their formation (Soker 2001).
 (2) The ionization structure near the central source changes
on a time scale of $5.5$ years (e.g., Smith {\it et al.} 2000),
which is taken to be the orbital period of the binary system
(Damineli 1986; Damineli {\it et al.} 2000).

\section{IONIZATION SHADOW}

 The proposed model for the formation of the strings is described in
section 2. We now derive the conditions for the model to work.
 Most of the basic physics used below can be found, e.g.,
in Canto {\it et al.} (1998).
 Let the proton and electron number densities in the strings'
surroundings be $n_p$ and $n_e$, respectively, and let $n_H$
be neutral hydrogen number density in the neutral shadow.
 At phase 3 of the model (see section 2), the surrounding gas is
fully ionized and optically thin to ionzing photons.
 The recombination of the surrounding gas yields a diffuse ionizing flux
of $F_d=n_e n_p \alpha_1 l_r/4$, where $l_r$ is the size of the
recombining region, which we take to be of the order of the radius $r$,
and $\alpha_1$ is the recombination coefficient to the ground state
of hydrogen.
 The ionization front proceeds into the neutral shadow at a speed of
$c_i \simeq F_d/n_H \simeq F_d/n_p$.
 For the shadow to be compressed we require $c_i<c_s$, where
$c_s \simeq 15 \km \s^{-1}$ is the sound speed of the ionized
surrounding gas.
Therefore, our condition for the compression of a long dense tail
in the shadow reads $15 {\km \s^{-1}} > n_e \alpha_1 r /4$, or
\begin{equation}
n_e \lesssim  200 \left( {{r}\over{ 2 \times 10^{17} \cm}} \right)^{-1} \cm^{-1},
\end{equation}
where for the distance we took the distance of the parts of the string
closest to the central star 75 years ago,
$r \simeq 2 \times 10^{17} \cm$ (considering the inclination
of the strings; see WDC).
 The electron density is a function of the mass loss rate, velocity,
and distance to the central star, hence the condition above can be
written as
\begin{equation}
\dot M_s  \lesssim 1.5 \times 10^{-4}
\left( {{r}\over{2 \times 10^{17} \cm}} \right) 
\left( {{v}\over{500 \km \s^{-1}}} \right) M_\odot \yr^{-1},
\end{equation}
where $\dot M_s/4\pi$ is the mass loss rate per unit solid angle
of the mass loss episode that formed the strings.

 The condition on the mass loss rate may seem very stringent, 
since the average mass loss rate during the great
eruption was much higher, $\dot M_{GE} \sim 0.1 M_\odot \yr^{-1}$.
 But noting the following we argue that this requirement is quite
reasonable.
 First, the strings are located near the long axis of the nebula$-$the
polar directions$-$and outside the dense part of the nebula.
 In the main nebula the mass loss rate per unit solid angle along
the polar directions is lower than near the equatorial plane, and
the mass loss rate which formed the regions outside the main
nebulae$-$the Homunculus$-$were much lower.
 Therefore, it is likely that the mass loss rate that formed the
strings was relatively low.
 Second, WDC show that the PN NGC 6543 contains strings which are
similar, in their shape, location, and relative brightness,
to those in $\eta$ Car.
 The maximum mass loss rates from asymptotic giant
branch progenitors of PNe are
$\dot M_{\rm max} \sim 10^{-4} M_\odot \yr^{-1}$.
  NGC 6543 does not contain a dense equatorial flow, and the mass
loss rate of its progenitor was probably lower, say 
$<3 \times 10^{-5} M_\odot \yr^{-1}$.
 The strings of NGC 6543, which are much fainter than the shell,
were formed from a much lower mass loss rate,
$\lesssim 3 \times 10^{-6} M_\odot \yr^{-1}$.
  We conclude that condition (9) for the formation of the strings
of $\eta$ Car by ionization shadow is reasonable.

 The immediately surrounding gas enters the shadow while
compressing the material in the shadow. 
 The recombining time of this material is
$\tau_{\rm rec} \simeq 10^3 (n_e/100 \cm^{-3})^{-1} \yr$, which is
longer than any other relevant time scale.
 However, the neutral material in the shadow is compressed to
an order of magnitude of larger densities, and after being ionized,
and before re-expanding, the recombination time is comparable to
the re-expansion time $\sim 100 \yr$.
 The recombination of the dense material of the strings makes their
formation process more efficient, since their ionization time
by the diffuse ionizing radiation takes somewhat longer.

 The typical width of the filaments is $\sim 0.002 \pc$
(Weis 2001).
 If the material was compressed to a density $\sim 10$ times higher,
the initial width of the material in the long strings
was $\sim 3$ times larger, i.e. $\sim 0.006 \pc$.
 As discussed above, the closest to the central star parts of the 
strings were at $r \simeq 0.06 \pc$ when the ionization of their 
surroundings started according to the proposed model.
 This means that the typical size (diameter if spherical) of
the shadowing clumps was
$D_{\rm clump} \sim 0.1 r \sim 2 \times 10^{16} \cm$.
A reasonable size for clumps formed by instabilities,
e.g., from winds collision (Dwarkadas \& Balick 1998).
  Narrower filaments, if formed, are short lived.
 First, the shadowing clump is small and it will evaporate
in a short time (assuming it is more or less spherical).
  Second, even if a compressed tail is formed, it will be ionized
and re-expand in a short time.
 Wider strings can in principle be formed. 
The question is whether dense clumps of such larger sizes exist.
 Since the formation process of the dense clumps is beyond the
scope of the present paper, we will not comment on that further.

\section{SUMMARY}

We proposed a model to explain the formation of the long and
narrow strings of $\eta$ Car, which are located along the symmetry
axis outside the Homunculus.
 Five such strings were identified by WDC in $\eta$ Carinae, and
similar strings exist in the PN NGC 6543 (WDC).
 The Hubble-type expansion law of the strings was beyond the scope of
the present paper, and we assumed that the material in the strings
had the Hubble-type expansion law before it was compressed. 
 The model assumes that dense clumps near the symmetry axis 
form an ionization shadow behind them. 
 The surrounding gas is ionized first, becomes much hotter, and
compresses the gas in the shadow. 
 This leads to the formation of a radial, dense, long, and narrow
region, i.e., a string. 
 Later the neutral material in the strings is ionized, and becomes
brighter.
Still later it re-expands, and we predicted that in
$\sim 100-200 \yrs$ the string would fade.

 We also showed that the companion, if it exists, may play a
significant role in the ionization of the strings' surroundings.
 The condition for the model to work is that the ionization front,
due to the diffuse ionizing recombination radiation of the 
surrounding gas, proceeds into the shadow at a velocity slower
than the compression speed, which is about the sound speed.
 From this we obtain a condition on the density of the strings'
surroundings (electron density in equation 8), or on the mass
loss rate of the mass loss episode that formed the strings (eq. 9).
 The mass loss rate is much below the average mass loss rate
during the Great Eruption of 1850, but we argue that is
is quite reasonable along and near the major axis and for
material outside the Homunculus.

\bigskip

{\bf ACKNOWLEDGMENTS:}
 This research was supported in part by grants from the
US-Israel Binational Science Foundation.


\end{document}